\documentclass[10pt, a4paper, twocolumn, aps, pra, showpacs, longbibliography, nofootinbib,superscriptaddress]{revtex4-1}
\pdfoutput=1
\usepackage[utf8x]{inputenc}
\usepackage{ucs}
\usepackage{amsmath}
\usepackage{amsfonts}
\usepackage{amssymb}
\usepackage{makeidx}
\usepackage{cellspace,booktabs}
\usepackage{natbib}
\usepackage{lipsum}
\usepackage{bm}
\usepackage{bbm}
\usepackage{relsize}

\usepackage{hyperref}
\hypersetup{colorlinks = true, linkcolor=magenta,citecolor=blue, urlcolor=magenta, bookmarksnumbered =  true}

\usepackage{color}
\definecolor{light-gray}{gray}{0.55}

\usepackage{microtype}

\usepackage{graphicx}

\newcommand{\ket}[1]{ \lvert #1 \rangle}
\newcommand{\braket}[2]{\langle #1 \vert #2 \rangle }

\begin{document}

\begin{abstract}
Large-scale quantum devices provide insights beyond the reach of classical simulations. However, for a reliable and verifiable quantum simulation, the building blocks of the quantum device require exquisite benchmarking. This benchmarking of large scale dynamical quantum systems represents a major challenge due to lack of efficient tools for their simulation. Here, we present a scalable algorithm based on neural networks for Hamiltonian tomography in out-of-equilibrium quantum systems. We illustrate our approach using a model for a forefront quantum simulation platform: ultracold atoms in optical lattices. Specifically, we show that our algorithm is able to reconstruct the Hamiltonian of an arbitrary size quasi-1D bosonic system using an accessible amount of experimental measurements. We are able to significantly increase the previously known parameter precision.
\end{abstract}

\date{\today}
\author{Agnes Valenti}
\thanks{These two authors contributed equally.}
\affiliation{Institute for Theoretical Physics, ETH Zurich, CH-8093, Switzerland}
\author{Guliuxin Jin}
\thanks{These two authors contributed equally.}
\affiliation{Kavli Institute of Nanoscience, Delft University of Technology, Delft, the Netherlands}
\affiliation{Institute for Theoretical Physics, ETH Zurich, CH-8093, Switzerland}
\author{Julian Léonard}
\affiliation{Department of Physics, Harvard University, Cambridge, Massachusetts 02138, USA}
\author{Sebastian D. Huber}
\affiliation{Institute for Theoretical Physics, ETH Zurich, CH-8093, Switzerland}
\author{Eliska Greplova}
\affiliation{Kavli Institute of Nanoscience, Delft University of Technology, Delft, the Netherlands}
\affiliation{Institute for Theoretical Physics, ETH Zurich, CH-8093, Switzerland}

\title{Scalable Hamiltonian learning for large-scale out-of-equilibrium quantum dynamics}

\maketitle

\section{Introduction}
Quantum simulators are at the forefront of quantum technologies and allow for simulation of specific quantum physics models using direct control and manipulation of existing quantum systems. A quantum simulation is generally tailored towards studying a specific type of phenomena typically chosen to be hard to simulate numerically on a classical computer. Quantum simulations have been successfully implemented in a range of platforms: trapped ions~\cite{blatt2012quantum, martinez2016real, zhang2017observation, zhu2020generation}, superconducting qubits~\cite{roushan2017chiral, king2018observation, ma2019dissipatively, chiaro2019}, semiconductor quantum dots~\cite{hensgens2017quantum}, and ultracold atoms~\cite{gross2017quantum, bernien2017probing, gorg2018enhancement, rispoli2019quantum, lukin2019probing, yang2020observation, ebadi2020quantum, wintersperger2020realization}. Many of these currently available experimental systems are reaching sizes that are prohibitive for their classical exact simulation~\cite{zhang2017observation, bernien2017probing}. This fact raises a challenge of the \emph{verification of quantum simulators}: While it is true that quantum simulators are able to address the issue of exploring physics that is intractable otherwise, our experimental control has intrinsic precision limits that lead to both errors and finite precision of quantum simulation. We therefore need to develop tools to verify the performance of quantum simulations to ensure that the correct physics is implemented.

Recent progress has been made in reconstructing generic local Hamiltonians from measurements on single eigenstates and steady states of closed and open systems~\cite{garrison2018does,qi2019determining,chertkov2018computational,bairey2019learning,bairey2020learning,evans2019scalable,cao2020supervised} as well as in the development of more specialized approaches tailored to concrete models~\cite{schirmer2004experimental, valenti2019hamiltonian, greplova2017quantum,devitt2006scheme,cole2005identifying,cole2006precision,burgarth2009coupling,av2020direct}.
An area of research that poses specific challenges for numerical methods and thus Hamiltonian tomography is out-of-equilibrium physics~\cite{eisert2015quantum,li2020hamiltonian,wiebe2014hamiltonian}. While many ground states and steady states of quantum systems can be readily captured by a range of approximate methods~\cite{foulkes2001quantum, carleo2017solving, cirac2020matrix}, out-of-equilibrium behavior of quantum systems proved to be significantly more challenging. Indeed, the difficulty in classically simulating out-of-equilibrium dynamics provide a readily available path towards quantum computational advantage~\cite{bermejo2018architectures, haferkamp2020closing}.

In the present work, we introduce a scalable method to learn the Hamiltonian governing out-of-equilibrium quantum simulations. We develop a neural-network based Hamiltonian learning technique that enables the reconstruction of the dynamics of large scale quantum system from experimentally accessible measurements with ultra-high precision at low computational cost. We exemplify our approach by learning Hamiltonian parameters governing the dynamics of quasi-1D quenched ultracold bosons in an optical lattice.


This manuscript is organized as follows. In Section~\ref{sec:Algo} we present our algorithm and exemplify its practical implementation on a quasi-1D out-of-equilibrium bosonic system. In Section~\ref{sec:Results} we present results of the Hamiltonian reconstruction for experimentally relevant parameter regime. In Section~\ref{sec:Bayes} we compare to a minimal Bayesian estimation benchmark in order to introduce context for the obtained estimation errors. In Section~\ref{sec:Scaling} we address how to experimentally scale our approach to arbitrary lattice sizes. Finally, in Section~\ref{sec:Conc} we present discussion and outlook of our results.

\section{Hamiltonian Learning}
\label{sec:Algo}
The central goal of our method is to reconstruct the Hamiltonian governing the dynamics of a quantum system, or, in other words, to find a set of parameters required to fully characterize the dynamics of the system under consideration. Additionally, we wish to reconstruct the Hamiltonian with a maximum possible accuracy from a practically accesible amount of experimental measurements. 
Our protocol relies on post-processing the measurement outcomes, and using the so-obtained accessible and efficient representation of the relevant information to train and evaluate neural networks for the parameter estimation.

Many practically and experimentally relevant Hamiltonians have a local structure, i.e. they are sums of terms that only act on sites with a finite separation. This structure does of course not prevent long range quantum correlations to arise in the system, but it largely simplifies their verification: Since our goal is to reconstruct Hamiltonians, not wave-functions, we may take advantage of this local structure and reconstruct the Hamiltonians from studying the physics of a subsystem of the experimentally relevant system. We show below that Hilbert spaces of these local subsystems become nearly or completely numerically manageable. Additionally, while the associated parameter space for the Hamiltonian of interest is not tractable, we show how to effectively approximate it in a scalable manner. 

We will explain our method on a concrete example of out-of-equilibrium quantum simulations with quasi one-dimensional  bosons in an optical lattice. The neutral bosonic atoms are trapped in an optical lattice formed by laser beams through the coupling of the dipole moment of the atom with an incoming light~\cite{jaksch1998cold, Bloch2008manybody}. This system has recently emerged as an excellent platform to explore out-of-equilibrium physics~\cite{rispoli2019quantum, lukin2019probing}.

The physics of trapped bosons in an optical lattice is described by the following  Bose-Hubbard Hamiltonian ~\cite{gersch1963quantum, jaksch1998cold}
\begin{equation}\label{BHhamiltonian_2}
	H_{\scriptscriptstyle\rm BH} = - \sum_{\langle ij \rangle}J_{ij} \hat{a}^{\dagger}_{i} \hat{a}_j
	+  \sum_i  \frac{U_i}{2} \hat{a}^{\dagger}_{i}\hat{a}_{i}(\hat{a}^{\dagger}_{i}\hat{a}_{i}-1)
	- \sum_i \mu_i  \hat{a}^{\dagger}_{i}\hat{a}_{i},
	\end{equation}
where the first term corresponds to particle hopping between neighboring sites (higher order hopping is suppressed) and the coefficients $J_{ij}$ correspond to the hopping amplitudes between site $i$ and site $j$. The second term corresponds to on-site interaction and the coefficients $U_i$ denote the strength of this interaction (in our case repulsion). The third term represents an on-site energy $\mu_i$. We are now tasked with determining the most probable values of the parameters $\theta=\{J_{ij}, U_i, \mu_i\}$ conditioned on the outcome of obtained measurements.

\begin{figure}
\centering
\includegraphics[width=\linewidth]{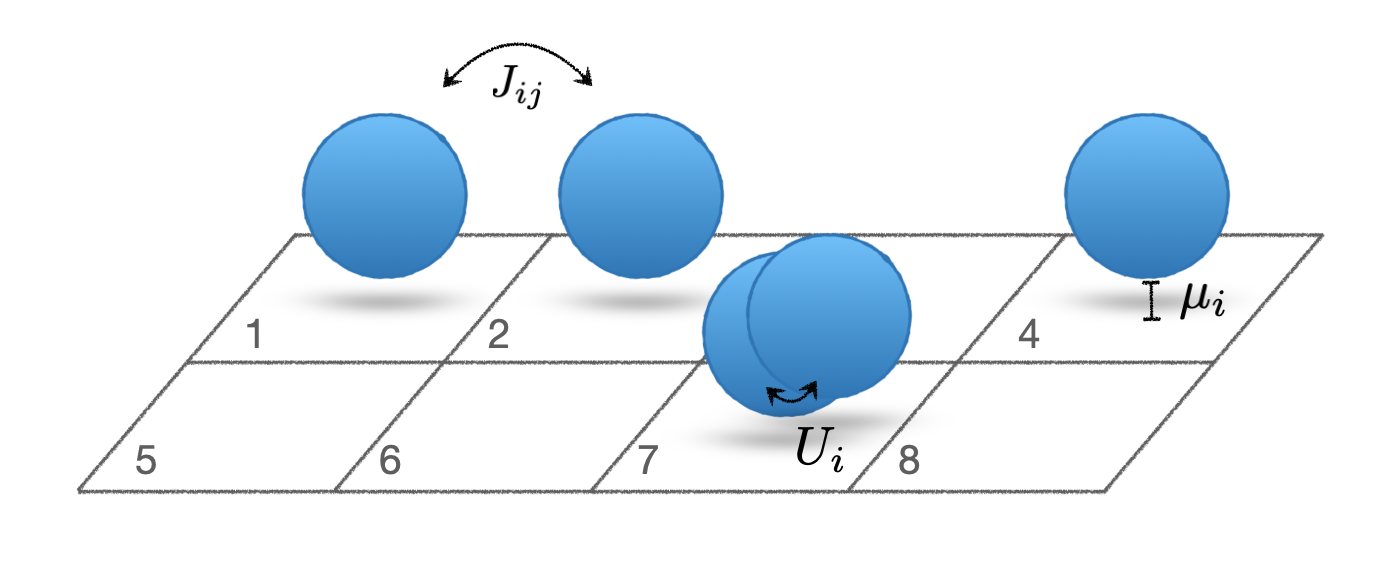}
\caption{Schematic plot of a subsystem of a quasi-1D optical lattice with bosons. Each square corresponds to a lattice site with the system parameters indicated. This system constitutes the building block of our estimation algorithm.}
\label{fig:intro}
\end{figure}

Since the Hamiltonian in Eq.~\eqref{BHhamiltonian_2} has a local structure where the operators at most act on a pair of neighboring sites, we consider whether we can reconstruct all the required parameters by measurements on a smaller, more tractable subsystem. Additionally, we expect the parameters to be reconstructed with higher precision on a smaller subsystem, as the amount of parameters dictating the system's behaviour and thus the measurement outcomes is reduced. In an experiment, it is possible to create smaller subsystems by locally blocking hopping between specific sites and thus isolating a number of selected particles into the pre-selected area of the lattice~\cite{aidelsburger2013realization}. While this sub-division might influence the Hamiltonian in the vicinity of the imposed borders, the parameters $J_{ij}$, $U_i$, and $\mu_i$, sufficiently distant from the border, will reflect their values in the extended system. In the example of quasi-1D lattice of size $ 2 \times L$ with $ L $ being the system length, let us consider a $2\times 4$ sites subsystem as shown in Fig.~\ref{fig:intro}. For $4$ bosons, the Hilbert space size is $330$ and thus the system is tractable via exact diagonalization. We use this tractable system of $8$ wells with $4$ particles as the basic building block for the training of our algorithm.

We consider the following out-of-equilibrium experimental sequence: the system is initialized in an easily prepared state with a trivial Hamiltonian followed by a quench of the Hamiltonian, i.e. a rapid change of the system parameters. After a time evolution for a time $T$ we perform a set of measurements on the system. While the specific measurement performed depends on the specific experiment, in the case of cold atoms experiments we have access to quantum gas microscopy~\cite{bakr2009quantum, sherson2010single} that allows for direct visualization of the occupation number in each well of the optical lattice. In other words, we have the ability to project onto the number states of the system after each time-evolution and therefore perform effective Monte-Carlo like sampling of the resulting quantum state. From these measurements we want to infer the Hamiltonian governing the time evolution as precisely as possible. Dependencies regarding the choice of initial state and time evolution time $T$ are discussed in Appendix~\ref{app:NN}.

In the following we explain how to design a computationally efficient machine learning algorithm for the Hamiltonian reconstruction. Additionally, we use the well-established, but computationally costly, Bayesian parameter estimation~\cite{bernardo2009bayesian, wiseman2009quantum} as a benchmark to assess the quality of the presented neural-network based Hamiltonian tomography process. Neural networks are, in general, capable of approximating an arbitrary function and once the network is trained it is possible to evaluate it on new input at low computational cost. 
Recent results suggest that classical machine learning is in general powerful in the analysis of quantum systems~\cite{torlai2019integrating, greplova2020unsupervised, palmieri2020experimental, zhang2020interpreting, bohrdt2020analyzing}.
Our goal is to approximate the map from post-processed experimental measurements, $M$, to the set of parameters that fully specify the Hamiltonian.
We note here, that a uniform chemical potential only yields a global phase factor and is thus not detectable within the closed system with a constant number of atoms. Instead, we reconstruct the fluctuations of the chemical potentials by considering the quantities $\mu_{diff,i}:=\mu_i-\mu_1$, $i \geq 2$. Thus, we aim to estimate the set of in total 25 parameters $\{J_{ij}, U_i, \mu_{diff,i}\}$ fully characterizing the system's dynamics.

An additional restriction that we are facing is the number of projective quantum gas microscope measurements (snapshots) that can be collected during the single experimental run. A single projection measurement in the number basis contains very little information both about the quantum state as well as the Hamiltonian parameters. On the other hand, if we had access to an infinite amount of projection measurements we could reconstruct the output distribution perfectly. These considerations introduce an additional optimization restriction to the problem we are trying to solve. We not only want to estimate the Hamiltonian with the maximum possible precision, but at the same time we need to be able to achieve that with as little data as possible. In the context of quantum gas microscopy, the practically achievable number of snapshots is on the order of $10^4$ \cite{rispoli2019quantum,lukin2019probing}.

Neural networks are trained using datasets of example data instances, the more examples the network sees the more general models we are able to build. Additionally, accurate results are more readily obtained when the function we wish to approximate has a low level of complexity. In order to reduce the complexity of the function mapping the measurement outcomes to the respective Hamiltonian parameters, we take two measures: Firstly we post-process the measurement outcomes such that the relevant physics are encoded in a more accessible manner. As the post-processing represents a compression of the measurement data, it yields the additional advantage of significantly reducing the training set size resulting in a tractable computational training cost.
Secondly, we factorize the problem by training a separate model for each parameter we wish to estimate.

The post-processing of the measurement data is based on calculating a set of relevant density correlators capturing the necessary information about the set of Bose-Hubbard parameters. Specifically we calculate the average occupation number $\langle n_i\rangle, i=1,\dots 8$ for each well. As the densities are in particular sensitive to the on-site repulsions and fluctuations of the chemical potentials, relevant information for the estimation of those parameters is compressed here.  Correlators of the density between any two $\langle n_in_j\rangle$, three $\langle n_in_jn_k\rangle$ and four $\langle n_in_jn_kn_l \rangle$ wells yield more complete information as they are also highly influenced by the values of the hopping amplitudes between wells. Additionally, we calculate the correlators $\langle n_i(n_i-1)\rangle$ in order to emphasize multiple occupancies. Given that we are using 4 particles in 8 wells, any higher order correlations are negligibly small. This construction results in $171$ correlation values that can be organized into a 1D vector $\vec n$, which presents a much more compact representation of the measurement dataset. We provide detailed information on the construction of the vector $\vec n$ in Appendix~\ref{app:NN}.

The factorization of the classification problem is performed as follows: instead of building a single model that predicts a vector of $25$ parameters for each measured dataset, we choose to build separate models for each parameter, i.e. $25$ models in total.

We use $25$ feed-forward neural networks trained using supervised learning. In all instances the input is the correlator vector $\vec n$ and the output is a single real number predicting the value of the given parameter. The networks consist of one input layer with 171 neurons (one neuron for each expectation value) and five fully connected hidden layers with 300, 400, 300, 150 and 100 neurons respectively.  The network is trained in a supervised manner using the squared difference between the predicted and correct parameter as cost function. This choice of cost function allows for accurate estimation of continuous parameters, representing a large advantage to discrete, more conventional methods as standard Bayesian parameter estimation models. We train each network using $150 500$ correlator vectors with parameters selected randomly from the confidence interval of the initial parameter estimation. We provide further detailed information on the architecture and training in Appendix~\ref{app:NN}.

\section{Results}
\label{sec:Results}

\begin{figure}[t]
\centering
\includegraphics[scale=1]{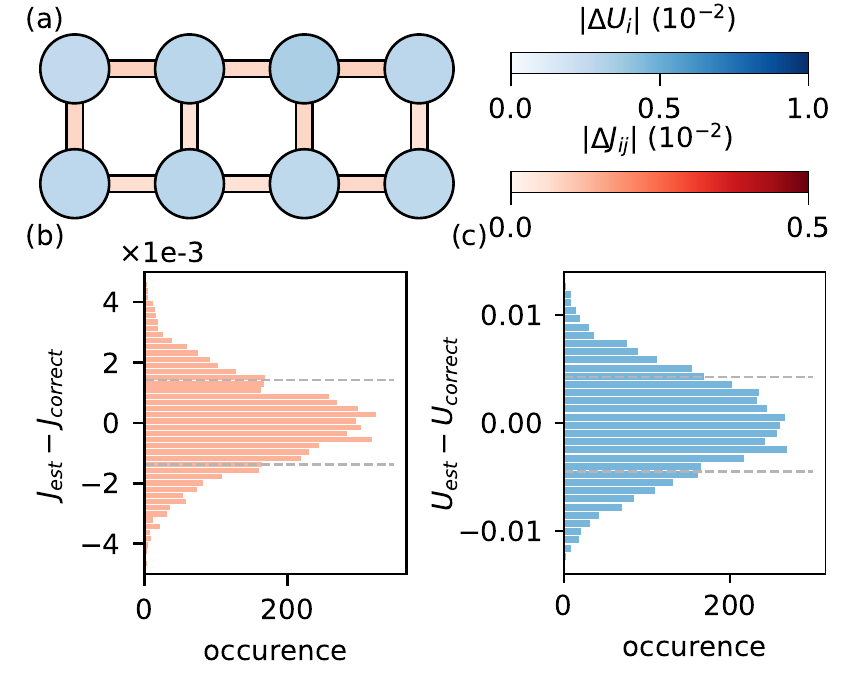}
\caption{Precision of the neural network estimation scheme for 2500 measurement snapshots: Panel (a) shows the size of the estimation errors regarding the respective positions of the on-site repulsion (blue circles) and the neighboring site hopping (orange connections). Panel (b) (Panel (c)) shows the error distribution independent of the spatial location for the hopping amplitudes (on-site repulsion) parameters over $500$ parameter configurations $\{J_{ij},U_i,\mu_i\}$, each evaluated using 2500 snapshots. Dashed grey lines denote the standard deviation of the error.}
\label{fig:ML-2500}
\end{figure}

\begin{figure}
\centering
\includegraphics[scale=1]{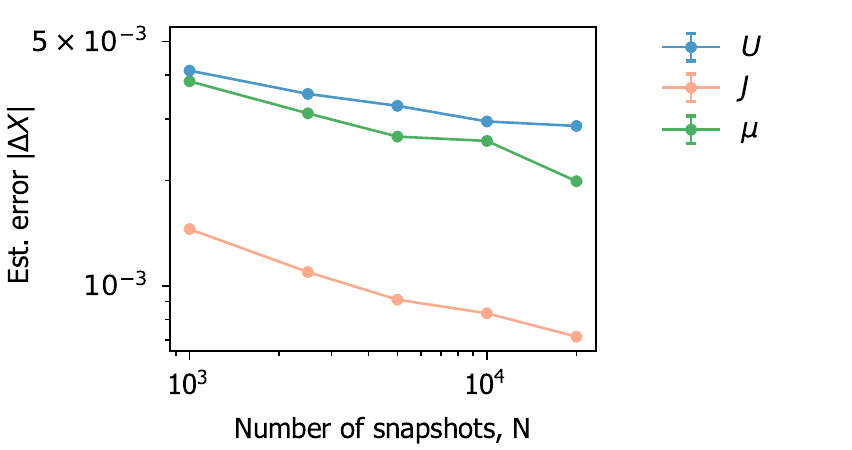}
\caption{Absolute estimation error as a function of number of measurement snapshots taken for neighbouring sites hopping amplitudes (orange), on-site repulsion (blue) and chemical potential differences (green). The average is taken over a test set of $500$ data sets, each set corresponding to a parameter configuration $\{J_{ij},U_i,\mu_i\}$.}
\label{fig:shots-number}
\end{figure}

We show the results of the neural network estimation in Fig.~\ref{fig:ML-2500} for the case of $2500$ measurement snapshots. Panel (a) shows the average errors for the hopping amplitudes $\Delta J_{ij}=|J_{ij}^{\textit{correct}}-J_{ij}^{\textit{estimated}}|$ and the on-site repulsion potentials $\Delta U_{i}=|U_{i}^{\textit{correct}}-U_{i}^{\textit{estimated}}|$ in their spatial location. The parameters are chosen such that on average we have $\langle J_{ij} \rangle = 1$, $\langle U_{i} \rangle = 2$ and $\langle \mu_{i} \rangle = 1$ and experimentally known within an interval of $1 \%$ ($\pm 0.5\%$ precision). The absolute uncertainty interval of the chemical potential differences $\mu_{diff,i}$ is by construction twice the uncertainty of the chemical potentials $\mu_{i}$. The blue circles in Fig.~\ref{fig:ML-2500} represent on-site repulsion errors and their red connections represent the inter-site hopping errors of the neural network estimation. The panels (b) and (c) show the distribution of the errors for hopping amplitudes and on-site repulsion, respectively. Grey dashed lines indicate the standard deviation of the distributions. 
The shown data has been averaged over test set consisting of $500$ measurement sets ($2500$ measurement snapshots each). We observe that the absolute error is around $0.1\times 10^{-2}$ for hopping amplitudes, $0.35\times 10^{-2}$ for the on-site repulsion and $0.3\times 10^{-2}$ for the chemical potential differences, therefore significantly improving over the prior precision for all parameters. We elaborate in Appendix~\ref{app:add} on the dependence of the estimated precision with respect to the experimentally known uncertainty.

As discussed above, the precision of the estimation depends on the number of measurement snapshots we have taken. In Fig.~\ref{fig:ML-2500} we have shown the precision for $2500$ measurements, i.e. a relatively low number. In Fig.~\ref{fig:shots-number} we show the scaling of the absolute error $\Delta X$ with $X\in\{J,U,\mu_{diff}\}$ as a function of measured snapshots with the error averaged over all hopping amplitudes, on-site repulsions and chemical potential differences.
More concretely, examining the errors $\Delta X$ for the hopping potentials, on-site repulsions and chemical potential differences, we see that in the most experimentally demanding case of $20 000$ snapshots, the averaged estimated absolute error is  $< 0.1\times 10^{-2}$, $< 0.3\times 10^{-2}$ and $< 0.25\times 10^{-2}$ for $J$, $U$ and $\mu_{diff}$, respectively, translating into a relative error of $< 0.1\%$ ($< 0.15\%$) for $J$ ($U$). When the number of snapshots is decreased, the error begins to slowly increase but remains low. The least demanding case of $1000$ snapshots per experimental realization results in an averaged estimated relative error of $< 0.15\%$ ($< 0.25\%$) for $J$ ($U$). The higher value of the absolute error for the on-site repulsion and the chemical potential differences can be explained with the known experimental precision. In particular, the absolute experimentally known precision interval is here set to be $0.02$ for both the on-site repulsions as well as the chemical potential differences, whereas it is of half the size for the hopping parameters. When rescaling the absolute errors through division with the experimental precision, the obtained errors (relative improvements of the known precision) are of similar magnitude for all parameters, see Appendix~\ref{app:add}.

\section{Bayesian estimation}
\label{sec:Bayes}

In order to assess the quality of the results provided by the neural network algorithm explained above we aim to establish a reliable benchmark.
In particular, we make a comparison with Bayesian parameter estimation. We note that a Bayesian classifier is provably optimal for classification problems~\cite{devroye2013probabilistic} and we expect similarly that Bayesian parameter estimation will make optimal use of the measurement data. However, as we explain below, as a consequence of the size of the parameter space, we cannot use the full potential of the Bayesian estimation due to the arising computational challenges. We can though still perform a Bayesian estimate for a slightly less complex problem in order to have a guiding threshold for the size of the estimation error.

Bayesian inference can be used to obtain the set of most likely parameters ${\theta}$ associated to measurement outcomes $M$. More specifically, this goal is achieved by connecting the probability distribution over the parameter space $\vec{\theta}$ with the experimental observations via Bayes' theorem. In particular, for a measurement outcome $M$ and the parameters $\theta$ we have
\begin{equation}\label{bayesmaintext}
P(\theta\mid M) = \frac{P(M\mid\theta)\cdot P(\theta)}{P(M)},
\end{equation}
where the likelihood $P(M\mid\theta)$ is the conditional probability to observe $M$ given the underlying parameter $\theta$, while $P(\theta\mid M)$ corresponds to the probability that the underlying parameter is $\theta$ given the measurement outcome $M$. $P(\theta)$ is the prior knowledge about the probability distribution of $\theta$ and $P(M)$ is the probability for $M$ and serves as normalization constant. Note that $P(M\mid\theta)$ can be obtained by analyzing the statistics of the experimental outcomes for a range of parameters $\theta$ and Bayes' rule~\eqref{bayesmaintext} allows us to determine the most probable $\theta$ given experimental outcomes. Let us now discuss how this applies to our example.

\begin{figure}
\centering
\includegraphics[scale=1]{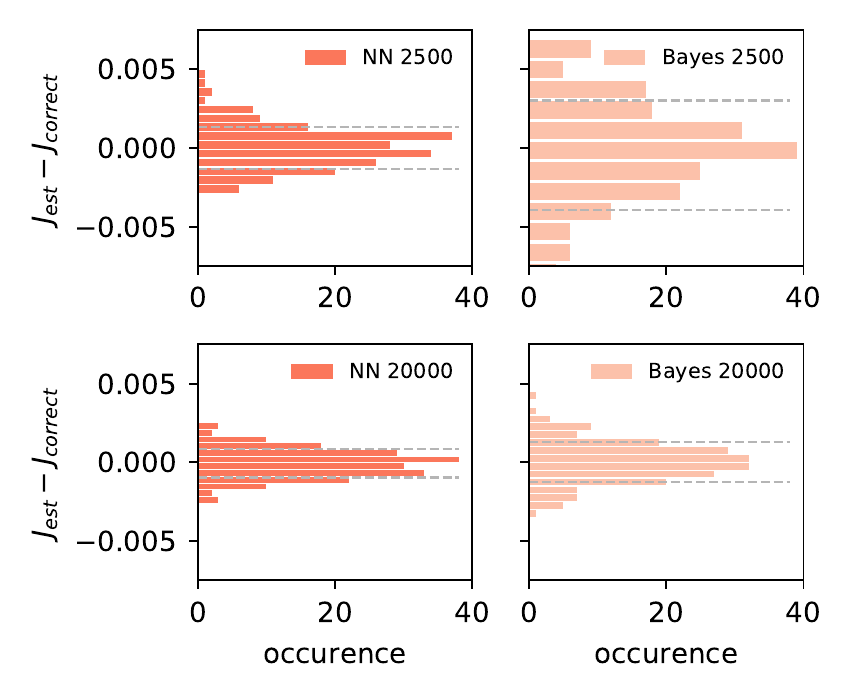}
\caption{Comparison of the performance of the neural network (left) and Bayesian method (right). The occurrences of the difference between estimated hopping and correct value are plotted as histogram for a set of 20 data sets each containing 2500 measurement snapshots (upper two panels) and 20 data sets each containing 20000 measurement snapshots (lower two panels). The standard deviation of the error is indicated via grey dashed lines.}\label{fig:comparison}
\end{figure}

The key simplification we have to make in comparison to the neural network is the discretization of the parameter space. When we train the machine learning model, we pick values of the Hamiltonian parameters randomly during the training from the desired confidence interval and the model then interpolates over the continuous interval. Bayes' rule, on the other hand, allows us to calculate likelihood for a set of candidate parameter values and choose the one that best fits with the measurement data. To find which parameter combination results in maximum likelihood, we need to evaluate the model on a finite grid.

We restrict ourselves to the estimation of hopping amplitudes and on-site interaction terms. On the basic building block subset of $2\times 4$ sites, this yields $18$ parameters to estimate. If we were to select the optimal values from only $5$ candidates per parameter, the number of parameter combinations to evaluate would be $5^{18} \approx 10^{12}$. As a consequence, we would need to perform an intractable number of these simulations to find out which specific parameter combination best fits to the experimental observation. 

Since we cannot test every point of the large parameter space probability distribution in a reasonable computational time, we will factorize this distribution as detailed in Appendix~\ref{app:bayes}. Essentially, we only estimate a small number of parameters at once, keep the rest fixed at the initial guess value and iterate until all estimated parameters have converged. The factorization reduces the number of quench simulations we need to perform to around $10^5$. We consider a grid of $13$ ($20$) candidates for each $J$ ($U$) and estimate all parameters with relative error $<0.25\%$ for the experimentally most challenging case of $20000$ snapshots. This number could be further explored by using a finer candidate grid. However, this estimate comes already at a very high computational cost: When implemented with highly parallelized code~\cite{ManyBodyDynLearning}, each instance of the Bayesian estimation takes approximately $24$ hours per measurement set when running the algorithm on $36$ cores. Additionally, there is no training and evaluation phase, the whole algorithm is re-run for each obtained measurement set - further adding to the computational demands over time.

We compare the performance of the neural-network based method with the results obtained via the Bayesian inference benchmark described here, see Fig.~\ref{fig:comparison}. We show the distribution of estimation errors over $20$ measurement sets for both the neural network and Bayesian estimation for $2500$ and $20000$ measurement snapshots. We observe that the machine learning algorithm shows a narrower error distribution and is, thus, firmly in the regime of errors comparable to Bayesian estimation while requiring only a fraction of the computational cost once trained. Additionally, we observe that with increasing number of measurement snapshots Bayesian estimation approaches the neural network error distribution. The neural network estimator does appear to suffer significantly less from error distribution broadening when decreasing number of measurements, thus outperforming the Bayesian estimation benchmark.


All code needed to recreate these results as well as a minimal Hamiltonian reconstruction demonstration can be found in \cite{ManyBodyDynLearning}.

\section{Scaling}
\label{sec:Scaling}

In the previous sections we have shown that we are able to estimate parameters with relative error $\sim  0.1\%$ ($\sim 0.15\%$) for $J$ ($U$) from as little as $2500$ measurement snapshots on a small subsystem of $4$ bosons in $8$ lattice sites of a quasi-1D bosonic lattice. The question remains how to experimentally scale the method up to larger systems while keeping the computational cost low. A first important observation is that the creation of the subsystems of the larger experimental system may introduce boundary effects. Specifically, we need to engineer the optical lattice to prevent any wave-function overlap between inside and outside the chosen subsystem. This action inevitably affects other parameters close to the introduced boundary. Using the site numbering introduced in Fig.~\ref{fig:intro}, the hopping amplitude between sites $1$ and $5$ (as well as $4$ and $8$) will likely be affected and it will lead to changes of effective on-site interaction in these sites. We therefore need to develop a strategy of 'shifting' our subsystem window such that we can take advantage of subsystem structure while capturing the whole system using as few measurements as possible.

\begin{figure}[t]
\centering
\includegraphics[width=\linewidth]{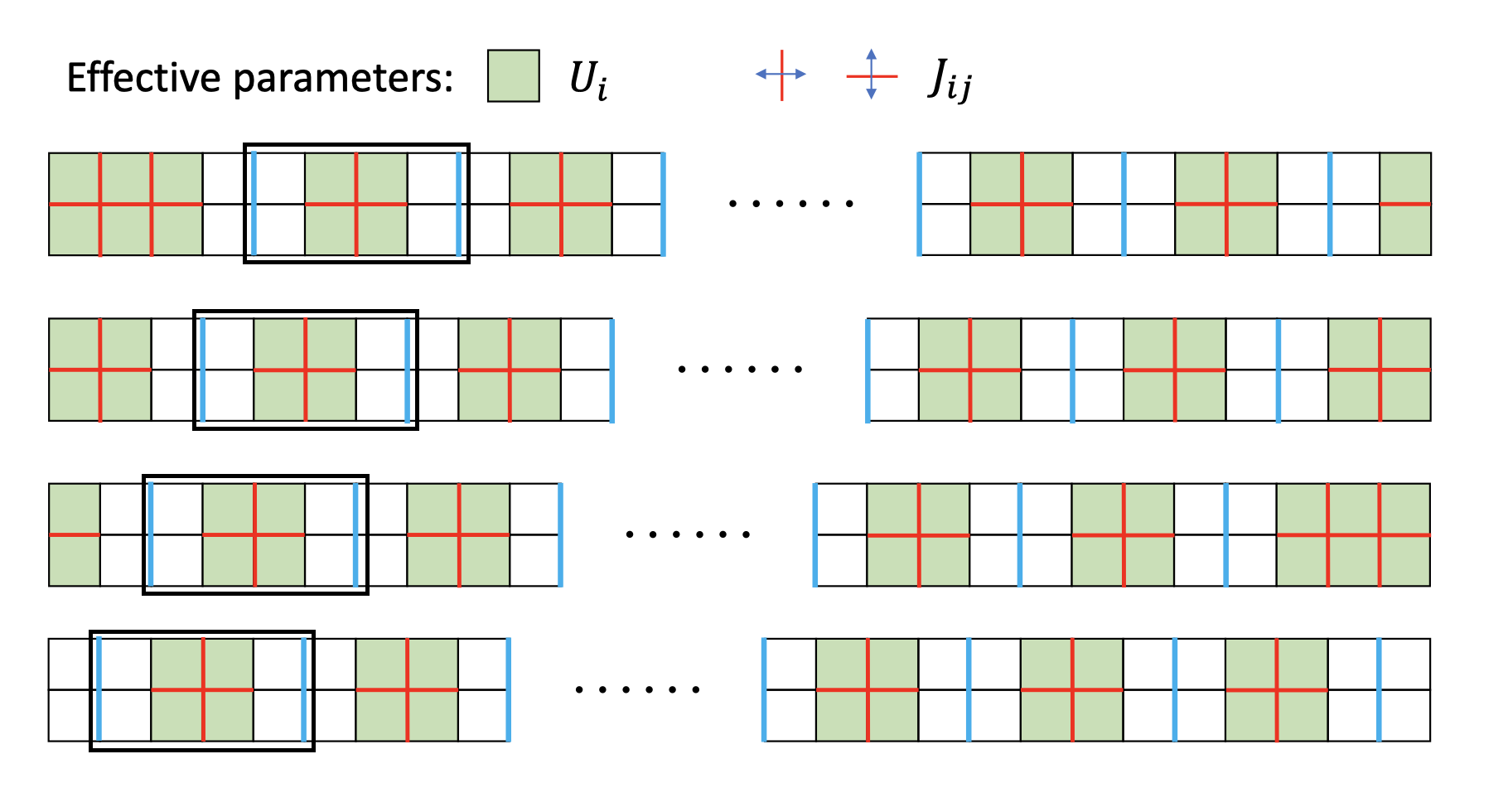}
\caption{Scaling of the protocol for the lattices of an arbitrary size. The black squares indicate a $2\times 4$ sub-system with the blue edge indicating a suppressed hopping term. The green boxes and the red edges correspond to the repulsion and hopping terms, respectively.}
\label{fig:scaling}
\end{figure}

We show this strategy in Fig.~\ref{fig:scaling}. Each row corresponds to one realization of the experiment, blue lines denote where the boundary should be risen and green squares and red lines correspond to interactions and hopping amplitudes respectively we want to infer. Our strategy is then the following: Step 1: raise boundaries in the system. Step 2: Take a projective measurement of the whole system and repeat for the desired number of snapshots. Step 3: Use the relevant part of the data to calculate expectation values for each subsystem prepared this way. Step 4: Shift the boundary one site to the left and repeat steps 2 through 4. As illustrated in Fig.~\ref{fig:scaling}, we only need to repeat this process four times since the fifth shift would lead to a system configuration that we have already measured. In practice this means that for our quasi-1D example, we can cover a lattice of arbitrary size with $4\times N$ snapshots. The final number of required measurements therefore depends on the desired precision as decided according to the information provided in Fig.~\ref{fig:shots-number}. For instance, considering $2500$ snapshots per boundary configuration we find that we can fully learn the Hamiltonian of the $2\times M$ quasi-1D lattice for any size of $M$ with $10 000$ experimental measurements.

\section{Discussion and Conclusions}
\label{sec:Conc}

Precise and robust Hamiltonian learning techniques that work reliably using only experimental data are key for verification of quantum simulators and manifestation of quantum advantage. The work presented here opens an avenue towards achieving just that and is immediately applicable to state-of-art quantum simulation experiments. In particular, we have developed a feed-forward neural network-based approach towards out-of-equilibrium Hamiltonian learning of large scale quantum simulators. By making use of the local structure of simulated Hamiltonians we build up a subsystem-based Hamiltonian learning method. We use a Bayesian estimation on a discrete parameter space as an estimation benchmark and then we design a series of neural networks that outperforms guiding Bayesian learning benchmark at a fraction of the computational cost. We have illustrated the effectiveness of our method on the Hamiltonian reconstruction of an out-of-equilibrium quasi-1D bosonic system in an optical lattice. Our specific example concerned the situation where the parameters can be relatively precisely estimated from first principles. Our method is then able to determine the parameters with a significantly higher precision than the initial guess. We show further examples of the results obtained using our algorithm for further parameter regimes and initial parameter guess precision in Appendix~\ref{app:add}.

When adapting our method to other systems the key challenge is the subsystem design. In the particular application explained here, we were able to select the subsystem small enough for us to be able to simulate the dynamics of the subsystem exactly. To increase the number of particles and fully capture a larger subsystem size, we may use a suitable approximate method to simulate the quantum dynamics~\cite{cirac2020matrix,gutierrez2019real,luo2020probabilistic} while approaching the factorization of the parameter space in an analogous fashion as described in this work.

\section*{Acknowledgements}
We are grateful for financial support from the Swiss National Science Foundation, the NCCR QSIT. This work has received funding from the European Research Council under grant agreement no. 771503.

\begin{appendix}

\section{Bayesian Estimation}
\label{app:bayes}

Bayesian inference is a statistical inference method for the calculation of the probability for a certain hypothesis based on available information \cite{casella2002statistical, feller2008introduction,degroot2012probability}. Specifically let us consider a machine, where a hypothesis corresponds to a parameter configuration $\theta$. Different machines specified by different $ \theta $ also differ in the probability distribution of their measurable outcome $ S $. The goal is to find which hypothesis (parameter configuration $\theta$) is most probable given the measurement $ S $. The relation of the internal parameters $\theta$ and accessible observables $ S $ is specified by Bayes' rule:
\begin{equation}\label{bayes}
P(\theta\mid S) = \frac{P(S\mid\theta)\cdot P(\theta)}{P(S)},
\end{equation}
where 
\begin{itemize}
	\item $ \theta $ can be any set of parameters specifying the machine. Our task is to determine which $ \theta $ is most probable according to the measured data $ S $.
	\item $ P(\theta) $ is the so-called prior probability. It contains our initial knowledge about the probability distribution of the parameters $\theta$ before taking into account the data $ S $.
	\item $ P(\theta\mid S) $ is the posterior probability specifying the conditional probability of a specific parameter configuration $\theta$ given the measured data $ S $.
	\item $ P(S\mid\theta) $ is the so-called likelihood: the probability of observing $ S $ given $ \theta $. 
	\item $ P(S) = \sum_{\theta} P(S\mid \theta)\cdot P(\theta)  $ is the normalization factor.
\end{itemize}
In our specific case the parameters $\theta$ correspond to all the unknown hopping amplitudes and on-site interactions $\theta=\{\vec J,\vec U\}$ and $S$ corresponds to quantum gas microscope measurement $M$ (snapshots). 

Note that in our case we do not only have a single observation $S$ but many thousands of them. For the case of many observations Bayes' rule generalizes as follows:
For $\{ S \} = (s_1,s_2,\ldots,s_n) $, where each observation $ s_i $ is independent and identically distributed (i.i.d), we can formulate the Bayes' theorem as \cite{gelman2013bayesian,choudhuri2005bayesian}:
\begin{equation}\label{multiplebayes}
P(\theta \mid \{ S \})={\frac {P(\{ S \} \mid \theta)\cdot P(\theta)}{P(\{ S \} )}},
\end{equation}
where the combined likelihood of a set of observations is given by the product of likelihoods for each individual observation,
\begin{equation}\label{combi}
P(\{ S \} \mid \theta)=\prod _{i}{P(s_{i}\mid \theta)}.
\end{equation}

In our case the observations $\{s_i\}$ correspond to all the possible Fock states one can obtain by occupational projections. The likelihood function $P(\{ S \} | \theta)$ is calculated according to Eq.~(\ref{combi}), given by the product of probabilities for measuring the corresponding Fock states 
\begin{equation}
\label{eq:productlikelihood}
P(\{ S \} \mid \theta)=\prod _{i}{P(s_{i}\mid \theta)} = \prod _{i}{P(\ket{i}\mid \theta)}.
\end{equation}
As for the prior, we consider a uniform distribution over the all the possible parameter configurations
\begin{equation}
    P(\theta) = \frac{1}{n_p},
\end{equation}
where $n_p$ is the number of candidate parameter configurations.

\begin{figure}
\centering
\includegraphics[scale=0.22]{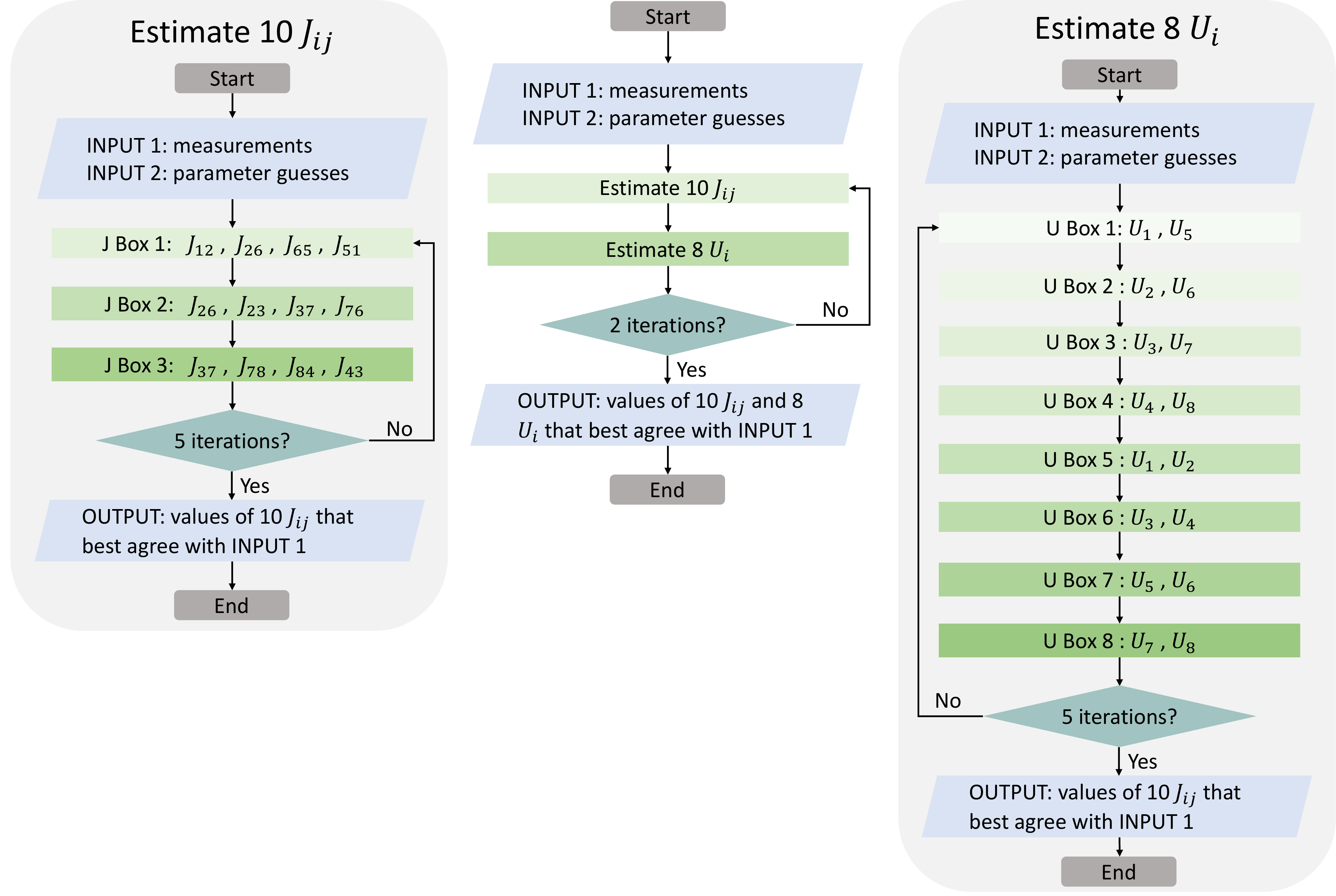}
\caption{Workflow of the Bayesian algorithm: 
due to the large parameter space, we evaluate in groups while keeping the rest constant according to the grouping scheme shown in the flowchart. One iteration corresponds to evaluating all 3 (8) sub-groups of $J$ ($U$). }
\label{fig:flowchart}
\end{figure}

Consider the system shown in Fig.~\ref{fig:intro}, we have $ n_p \sim 10^{18}$ when we assign 10 candidates for each $J$ and $U$. This parameter space size exceeds the capacity of any contemporary computer. Therefore we need to further factorize the parameter estimation process. We estimate a small group of parameters each time and update the values, iterating over all parameter groups. We begin with 10 hopping amplitudes $J$, then move on to 8 on-site repulsions $U$ and consequently we come back to re-estimate $J$ and $ U $. The hopping amplitudes $J$ and on-site repulsions $U$ are in addition factorized in smaller sub-groups accordingly to their spatial location. The details of this factorization process are shown in Fig.~\ref{fig:flowchart}. We prepare 8 groups of $U$ and 3 groups of $J$. 


We discretize the parameter space accordingly to the experimentally known uncertainty. In particular, for the experimental precision of $\pm 0.5 \%$, the correct values of the parameters lie within the intervals  $ [0.995,1.005]$ for $J$, $[1.99,2.01]$ for $U$ and $ [0.995,1.005]$ for $\mu$ and we choose $13$ parameter candidates for $J$ and $21$ candidates for $U$ as equally spaced within these intervals. During the estimation process, the simulations are performed using the mean value of the chemical potential within the known precision, i.e. $\mu=1$. As the correct value may differ, an error is introduced by this choice. The accuracy of the estimation process might increase by including the estimation of the chemical potential, we here restrict ourselves to the estimation of $J$ and $U$ due to the computational cost.

The iterative parameter estimation is performed as follows. The estimation is based on $N$ experimental snapshots which are re-used in every iteration. As we factorized the parameter estimation, we only estimate the parameters within a subgroup at a time, while keeping the rest of the parameters fixed. In particular, we initially set the values of the parameters outside of the current subgroup to the mean values within the experimental uncertainty (i.e. $J = 1.0$, $U=2.0$) and update them accordingly to their estimated values after each estimation.

We start the parameter estimation process by estimating the hopping amplitudes. Concretely, we subsequently estimate the $J_{ij}$ per subgroup ($3$ subgroups) and update the parameter values accordingly. Within each subgroup of the hopping amplitudes, we estimate $4$ hopping amplitudes at a time. Using $13$ parameter candidates for each hopping amplitude, this yields a total dimension of $13^4 = 28561 $ for the parameter space.
After the first estimation of all hopping amplitudes, the on-site repulsions are estimated by subgroup. As each subgroup here only contains two $U_i$ to be estimated, we can choose a finer grid ($21$ candidates), while keeping a feasible parameter space dimension of $21^2 = 441 $.
Estimating all on-site repulsions completes the first iteration, and the process is repeated. In total, we perform $5$ iterations of the estimation process.


In summary of this appendix section, we employ Bayes' rule in the following steps:
\begin{enumerate}
\item
Prepare 'experimental' data: We initialize the system in a selected Fock state, simulate unitary evolution under a Hamiltonian with randomly selected correct parameters, followed by a projective measurement. We repeat this process by the number of times that corresponds to the selected number of snapshots, $N$. We only keep the measurements $\{ S \}$ for the follow-up estimation.

\item
Simulated likelihood: We simulate the unitary evolution under each Hamiltonian specified by all the possible candidate configurations from the parameter space. We factorize this process as shown in Fig.~\ref{fig:flowchart}. For each candidate parameter configuration $ \theta $, we take the final state (after time evolution $T$) and calculate the overlap with all the Fock basis states respectively: 
$\{\braket{\Psi_{\textit{final}, \theta}}{i}\}$.
This yields the ingredients $\{ P(\ket{i}\mid \theta)\}$ we need to calculate the likelihood function as shown in Eq.~\eqref{eq:productlikelihood}.
\item
Select the most likely parameters: The parameter candidates $\theta$ for which the set 
$\{ P(\ket{i}\mid \theta)\}$ 
can maximize the likelihood \eqref{eq:productlikelihood} is most likely to generate the same physics observed in the experiment.

\item
Update: Update the current values of the parameters in the current group by the Bayesian estimated values.

\end{enumerate}

\begin{figure}
\centering
\includegraphics[scale=1]{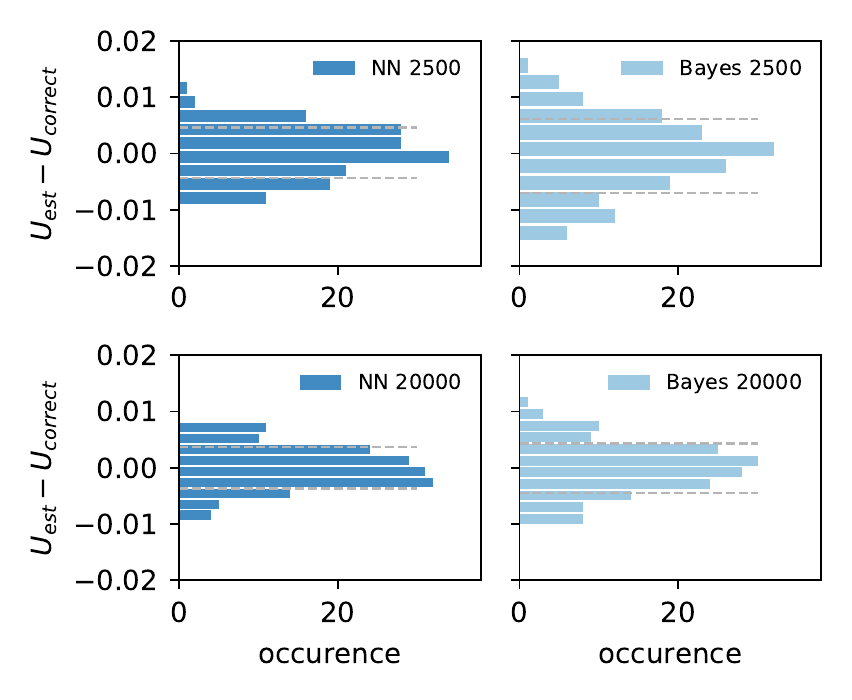}
\caption{Comparison of the performance of the Neural network (left) and Bayes' method (right). The occurrences of the difference between estimated on-site repulsion and correct value are plotted as histogram for a set of 20 data sets each containing 2500 measurement snapshots (upper two panels) and 20 data sets each containing 20000 measurement snapshots (lower two panels). The standard deviation of the error is indicated via grey dashed lines.}\label{fig:comparison_U}
\end{figure}

We have shown the comparison between the neural-network based estimation and the Bayesian inference in Fig.~\ref{fig:comparison} for the estimation of hopping amplitudes, where the neural network outperforms the Bayesian estimation. We show in addition the results for the comparison between the estimation of the on-site repulsions provided by the neural network and using Bayesian inference in Fig.~\ref{fig:comparison_U}. While the neural network still yields more accurate results, the difference is less striking compared to the hopping amplitude estimation in Fig.~\ref{fig:comparison}.

\section{Neural Networks}
\label{app:NN}

\begin{figure}
\centering
\includegraphics[scale=0.13]{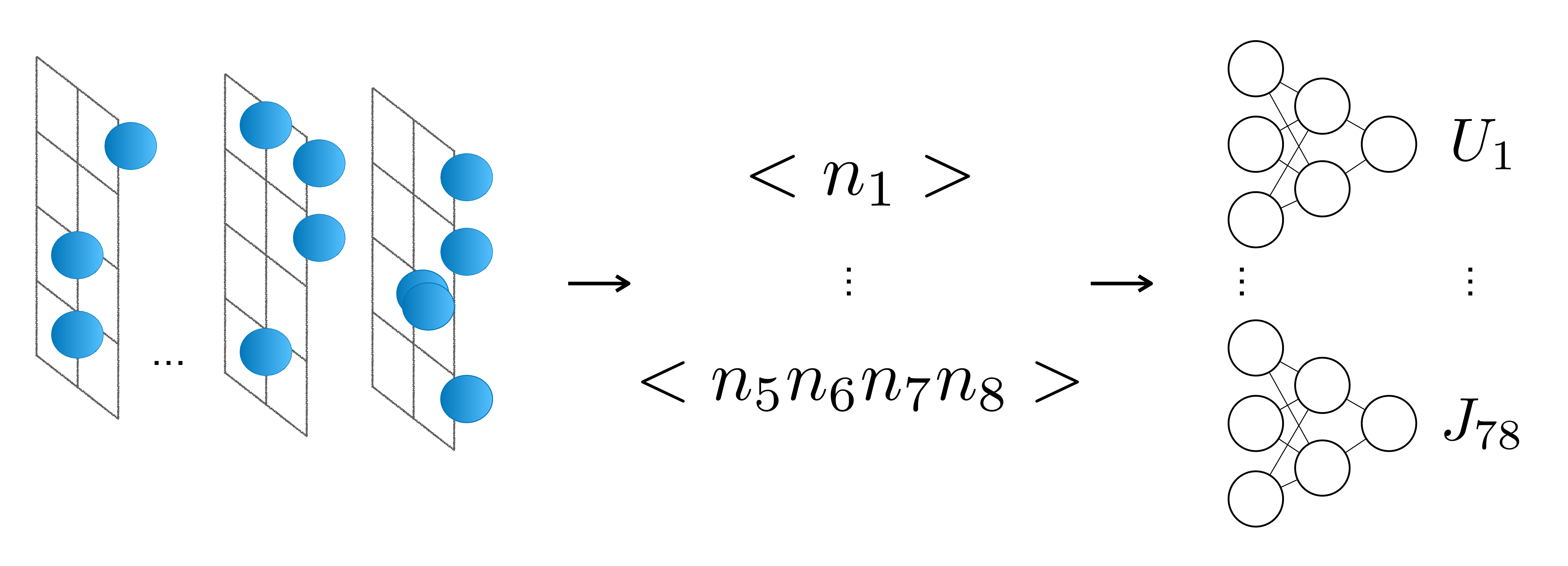}
\caption{Workflow of the neural network-based algorithm: 1. Measurement snapshots are experimentally taken or numerically simulated. 2. Compute the set of relevant correlators. 3. The set of correlators then serves as an input of $25$ feed-forward supervised networks, one for each parameter to be estimated.}
\label{fig:NN-scheme}
\end{figure}

In the following, we provide a detailed description of the neural-network based Hamiltonian reconstruction scheme.

As explained in the main text, the goal is to provide an estimation of the parameters of the Bose-Hubbard Hamiltonian Eq.~\eqref{BHhamiltonian_2}  based on atom density measurements after a time evolution of a prepared initial state. In particular, we choose an initial state with $4$ atoms, which constitutes the upper limit of bosons on a $2\times 4$ -lattice that is still feasible for our approach, as it is fast to simulate via exact diagonalization. We have found that the particular initial distribution of the $4$ bosons on the lattice plays a negligible role in the precision of the parameter estimation and can therefore be chosen arbitrarily, or as a distribution which is experimentally most feasible. 
The initial state used exemplary here contains an atom on every second site. Measurements are taken after a time evolution under the Bose-Hubbard Hamiltonian Eq.~\eqref{BHhamiltonian_2} of $T$, where we here choose as example $T=200/\langle J_{ij} \rangle$ to be experimentally realizable. Testing with smaller $T$ did not show any significant difference in precision, therefore the neural network may be trained for a specific experiment using the experimental most experimentally suitable time evolution $T$.

For each of the hopping amplitudes $J_{ij}$ and on-site repulsions $U_i$, a separate network is trained to estimate the parameter's value. The network takes as input post-processed measurement data and outputs the respective parameter value. In particular, from $N$ density snapshots conducted on the $2\times 4$-lattice in a first step a set of expectation values are calculated in order to compress the data.
More concretely, we calculate the following expectation values:
\begin{itemize}
    \item the single-site densities $\langle n_i \rangle$ with $n_i$ being the number of atoms on site $i$,
    \item the two-point density correlation functions $\langle n_i n_j \rangle$, $i<j$
    \item the three-point density correlation functions $\langle n_i n_j n_k \rangle$, $i<j<k$
    \item the four-point density correlation functions $\langle n_i n_j n_k n_l \rangle$, $i<j<k<l$
    \item  $\langle n_i (n_i-1) \rangle$ counting multiple occupancies on site $i$,
\end{itemize}
For $8$ lattice sites, this yields a total of $171$ expectation values. This number, and consequently the network input dimension, is independent of the number of conducted measurement snapshots.
For each parameter to be estimated, a network is trained taking as input $171$ expectation values calculated from the experimental measurements to output the respective parameter's value. The parameter estimation workflow is depicted in Fig.~\ref{fig:NN-scheme}.

\begin{figure}[t]
\centering
\includegraphics[scale=1]{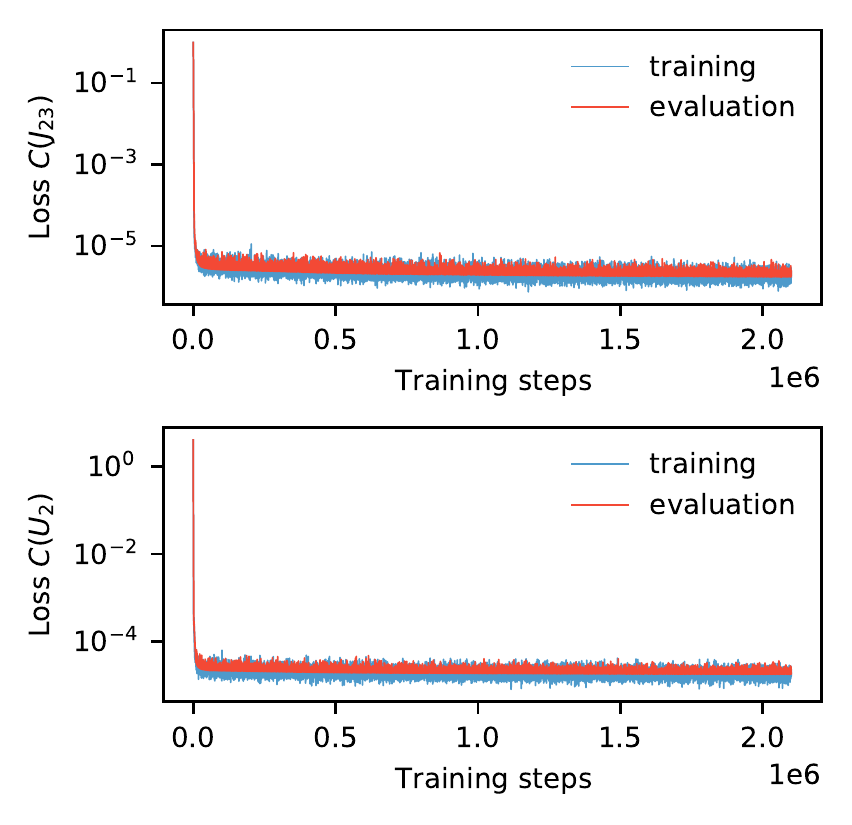}
\caption{Network training (blue) and evaluation loss (red) as a function of training steps for the network estimating the hopping parameter $J_{23}$ (upper panel) and the network estimating the on-site repulsion $U_2$ (lower panel).}
\label{fig:NNtraining}
\end{figure}

We use a fully connected neural network with one input layer with $171$ neurons (one neuron for each expectation value), $5$ hidden layers with $300$, $400$, $300$, $150$ and $100$ neurons and an output layer with one neuron. The value of the output neuron corresponds to the parameter to be estimated. The network is trained using supervised learning and the mean squared loss function $C(X)$ for continuous parameter estimation:
\begin{align}
C(X)=\frac{|X^{\rm pred}-X^{\rm label}|^2}{N_{\rm Batch}}.
\end{align}
Here, we use $X$ as placeholder for the parameter $J_{ij}$, $U_i$ or $\mu_{diff,i}$ to be estimated. $X^{\rm label}$ corresponds to the correct value whereas $X^{\rm pred}$ is the neural-network prediction. The Batch size during training is given by $N_{\rm Batch}$.

We create a training set of $150500$ training examples, each example consisting of the expectation values together with the respective parameter label. The time evolution and conducted measurements are simulated via exact diagonalization. For each of the training examples, the correct parameter values are chosen randomly within the interval of confidence, here $J_{ij}=1 \pm 0.005$, $U_{i}=2 \pm 0.01$ and $\mu_{i}=1 \pm 0.005$.
Using the learning rate $\eta=10^{-5}$ and a batch size of $N_{\rm Batch}=50$, the training is conducted for a total of $2.1\times 10^{6}$ training steps. The training and evaluation loss for the two networks estimating the parameters $J_{23}$ and $U_2$ using $2500$ measurement snapshots are shown in Fig.~\ref{fig:NNtraining}, where the evaluation set corresponds to a set of 500 examples not used for training.
The neural network was trained using the library Tensorflow \cite{abadi2016tensorflow}.

\section{Additional parameter regimes}
\label{app:add}

\begin{figure}
\centering
\includegraphics[scale=1]{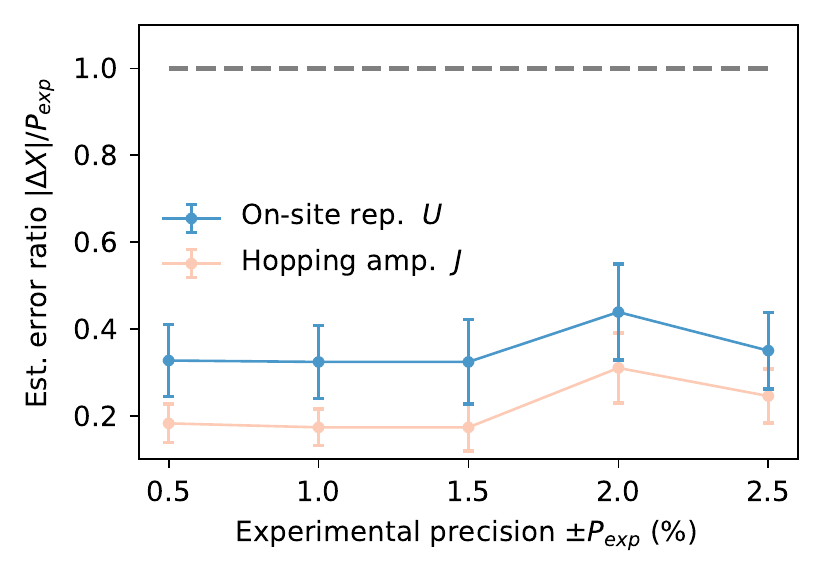}
\caption{The ratio of the parameter estimation error of the on-site repulsion $\Delta X=\Delta U$ and the hopping strengths $\Delta X=\Delta J$ divided by the experimental precision $P_{exp}$ as a function of $P_{exp}$. The grey dashed line indicates the ratio $P_{exp}/P_{exp}=1$ to emphasize the improvement with respect to the experimental precision. We use here the neural-network based parameter estimation with $5000$ snapshots.}
\label{fig:ML-interval}
\end{figure}

\begin{figure}
\centering
\includegraphics[scale=1]{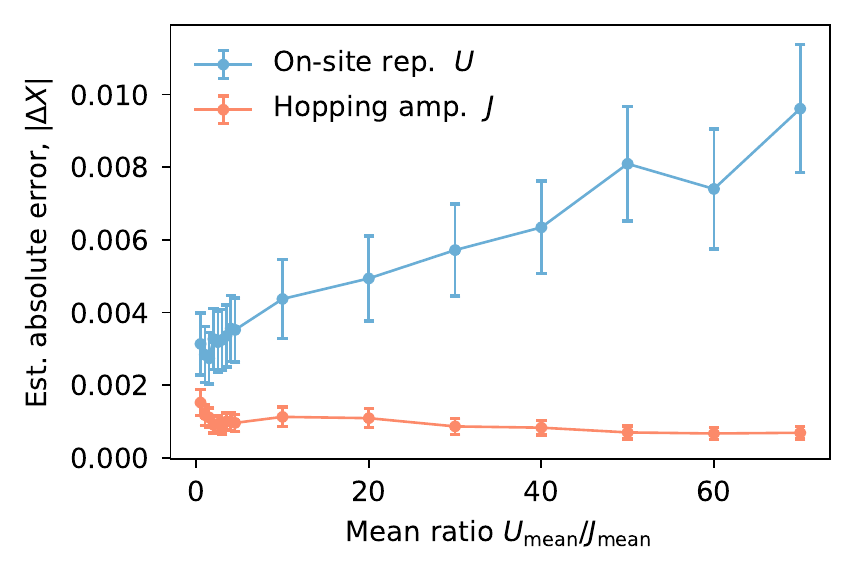}
\caption{The absolute parameter estimation errors of the on-site repulsion $\Delta X=\Delta U$ and the hopping strengths $\Delta X=\Delta J$ as a function of the ratio $U_{\rm mean}/J_{\rm mean}$. We use here the neural-network based parameter estimation with $5000$ snapshots.}
\label{fig:ML-shiftedU}
\end{figure}

In the following, we examine the dependence of the parameter estimation accuracy for varying parameter regimes.
We start by analyzing the dependence on the size of the parameter confidence interval, i.e. the previously known experimental precision. In the main text, all simulations are done with a previously known parameter precision of $1 \%$ ($\pm 0.5 \% $) and the neural network estimation improves this precision by half an order of magnitude. 

The initially known precision might vary from this choice. In particular, we consider parameter precisions up to $\pm 2.5 \%$ and examine the relative improvement in precision obtained after applying the neural network Hamiltonian reconstruction scheme. For each interval, we re-train neural networks with training data within the experimentally known precision. The relative improvement of the initial precision $P$, i.e. the ratio $\Delta X/P$ is shown in Fig.~\ref{fig:ML-interval} for $X=J,U$. Using $5000$ snapshots, we obtain a significant gain in precision which does not notably increase within the interval $P \in [0.5 \%, 2.5 \%]$.

We examine in addition, how the parameter estimation precision varies with respect to the ratio of the on-site repulsion $U$ to the hopping strength $J$. In particular, we vary the (average) strength $U_{\rm mean}=0.5, ... 70$ of the on-site repulsion $U_{i}=U_{\rm mean} \pm 0.01$ while keeping the hopping amplitudes fixed ($J_{\rm mean}=1.0$) in the interval $J_{ij}=J_{\rm mean} \pm 0.005$. The {\it absolute} precision of $U$ is unchanged for varying $U_{\rm mean}$ in order to ensure a meaningful comparison of the obtained parameter estimation precisions. The absolute parameter estimation errors
\begin{align}
    \Delta U=\frac{1}{8}\sum \limits_{i} |U_i^{pred}-U_i^{label}|, \\
    \Delta J=\frac{1}{10}\sum \limits_{\langle ij \rangle} |J_{ij}^{pred}-J_{ij}^{label}|
\end{align}
are plotted in Fig.~\ref{fig:ML-shiftedU} as a function of $U_{\rm mean}/J_{\rm mean}$ using $5000$ snapshots. Here, the sum in the second line runs over all nearest-neighbour bonds. 

Figure~\ref{fig:ML-shiftedU} shows an increasing absolute error in estimating the value of the on-site repulsion for increasing ratio $U_{\rm mean}/J_{\rm mean}$. At the same time, the estimation error of the hopping amplitudes slightly decreases for increasing $U_{\rm mean}/J_{\rm mean}$. We can understand this result by examining the interplay of the on-site repulsion and the hopping strengths. More specifically, the timescale of the atom movement between the different sites is set by the hopping strength $J_{\rm mean}$. The on-site repulsion might be roughly understood as influencing the 'direction' of hopping: The stronger the on-site repulsion, the more the probability of measuring a configuration with several atoms on a site is reduced. As our simulations take place with $4$ atoms on $8$ lattice sites, the configuration is sufficiently sparse such that double occupancies may be avoided. For very large ratio $U_{\rm mean}/J_{\rm mean}$, we the probability to measure a double occupancy becomes increasingly small and therefore unlikely to observe with a finite number of measurement snapshots. Small changes in the on-site repulsion strength are therefore expected to be increasingly difficult to detect, as the effect of the on-site repulsion manifests itself mainly in the occurence of multiple occupancies of lattice sites. This behaviour is shown in Fig.~\ref{fig:ML-shiftedU}.
A more dense configuration of atoms (i.e. a more than half filled lattice) might show a different behaviour, as the probability of mulitply occupancies is not suppressed in the same way compared to a sparse atom configuration.

\section{Estimating chemical potentials: details}
\label{app:mu}


\begin{figure}
\centering
\includegraphics[width=\linewidth]{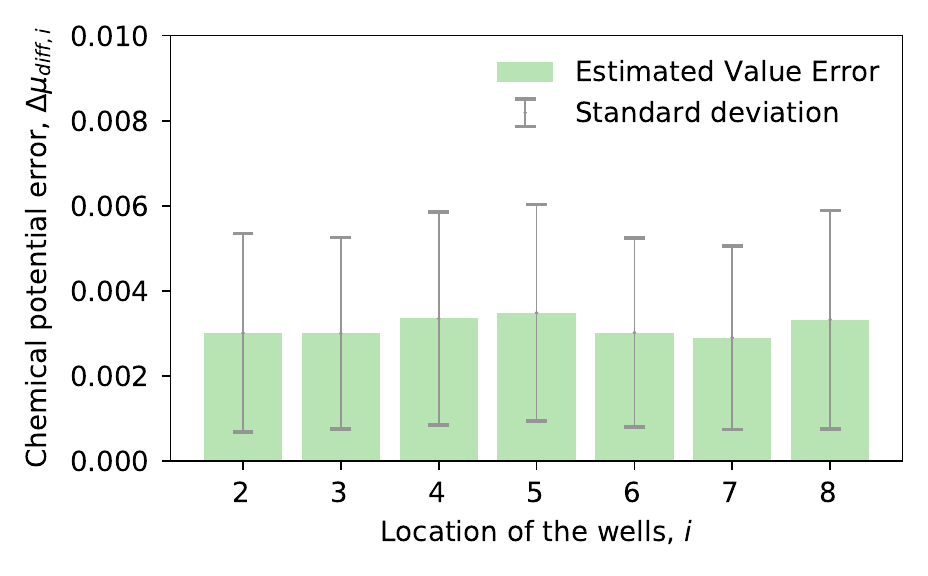}
\caption{Estimation error of the neural network estimation scheme for the chemical potential differences $\mu_{i,diff}=\mu_i-\mu_1$ for 2500 measurement snapshots as a function of the spatial location $i$, averaged over $500$ data sets. }
\label{fig:ML-mu_2500}
\end{figure}

\begin{figure}
\centering
\includegraphics[width=\linewidth]{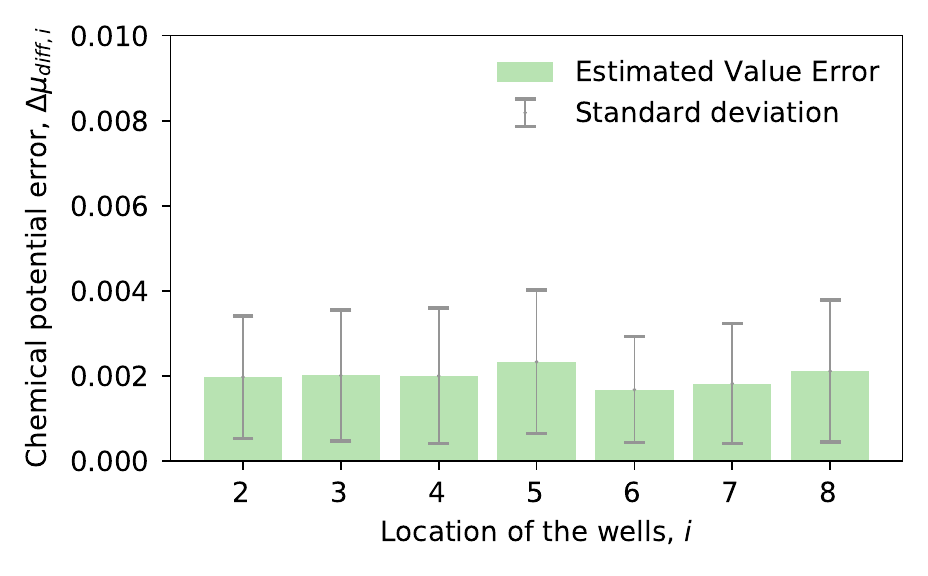}
\caption{Estimation error of the neural network estimation scheme for the chemical potential differences $\mu_{i,diff}=\mu_i-\mu_1$ for 20000 measurement snapshots as a function of the spatial location $i$, averaged over $500$ data sets. }
\label{fig:ML-20000}
\end{figure}

As we are considering only states with a constant total atom number (in particular, we use $N=4$ atoms), a uniform chemical potential $\mu$ yields only a global phase factor $e^{-i/\hbar \mu N}$. Thus, a uniform chemical potential does not affect the system's dynamics and is as a consequence not detectable within the closed system. Relevant and measurable effects are instead induced by a {\it non-uniform} chemical potential. We can quantify the deviations of the chemical potential by considering e.g. the differences $\mu_{diff,i}:=\mu_i-\mu_1$, $i \geq 2$. 

When estimating the parameters $\{J_{ij}, U_{i}, \mu_{diff,i}\}$, we vary all parameters within a percent, i.e. $\mu=1 \pm 0.005$
(and $J=1.0\pm 0.005 $, $U=2.0\pm 0.01 $). As a consequence, the differences $\mu_{diff,i}=\mu_i-\mu_1$ take the values $\mu_{diff,i}=0 \pm 0.01$. We show the obtained estimation error 
\begin{align}
\Delta \mu_{diff,i}=|\mu^{label}_{diff,i}-\mu^{est.}_{diff,i}|
\end{align}
 and their spatial dependence in Fig.~\ref{fig:ML-mu_2500} and Fig.~\ref{fig:ML-20000} for 2500 and 20000 measurement snapshots, respectively.

\end{appendix}

\bibliography{references.bib}

\end{document}